\documentclass[nohyperref]{article}

\usepackage{microtype}
\usepackage{graphicx}
\usepackage{subfigure}
\usepackage{booktabs} % for professional tables
\usepackage{microtype}
\usepackage{graphicx}
\usepackage{subfigure}
\usepackage{booktabs} % for professional tables
\usepackage{amsmath}
\usepackage{amsfonts}
\usepackage{mathtools}
\usepackage[]{amssymb}
\usepackage{bm}
\usepackage{rotating}
\usepackage{dutchcal}
\usepackage{xfrac}

\DeclareMathOperator{\ERLE}{ERLE}
\DeclareMathOperator{\PESQ}{PESQ}

\newif \ifwebcolor
\webcolortrue

\usepackage{hyperref}

\usepackage[accepted]{icml2022}

\usepackage{amsmath}
\usepackage{amssymb}
\usepackage{mathtools}
\usepackage{amsthm}

\usepackage[capitalize,noabbrev]{cleveref}

\theoremstyle{plain}

\theoremstyle{definition}

\theoremstyle{remark}

\usepackage[textsize=tiny]{todonotes}

\icmltitlerunning{Neural Harmonium}

\begin{document}

\twocolumn[
\icmltitle{Neural Harmonium: An Interpretable Deep Structure for Nonlinear Dynamic System Identification with Application to Audio Processing}



\icmlsetsymbol{equal}{*}

\begin{icmlauthorlist}
\icmlauthor{Karim Helwani}{to}
\icmlauthor{Erfan Soltanmohammadi}{to}%
\icmlauthor{Michael M. Goodwin}{to}
\end{icmlauthorlist}

\icmlaffiliation{to}{Amazon Web Services, Inc., Palo Alto, CA, USA}

\icmlcorrespondingauthor{Karim Helwani}{helwk@amazon.com}

\icmlkeywords{System identification, online processing, interpretable, nonlinear systems}

\vskip 0.3in
]



\printAffiliationsAndNotice{}  % leave blank if no need to mention equal contribution

\begin{abstract}
	Improving the interpretability of deep neural networks has recently gained increased attention, especially when the power of deep learning is leveraged to solve problems in physics. Interpretability helps us understand a model's ability to generalize and reveal its limitations. In this paper, we introduce a causal interpretable deep structure for modeling dynamic systems. Our proposed model makes use of the harmonic analysis by modeling the system in a time-frequency domain while maintaining high temporal and spectral resolution.
	Moreover, the model is built in an order recursive manner which allows for fast, robust, and exact second order optimization without the need for an explicit Hessian calculation. To circumvent the resulting high dimensionality of the building blocks of our system, a neural network is designed to identify the frequency interdependencies. The proposed model is illustrated and validated on nonlinear system identification problems as required for audio signal processing tasks.  Crowd-sourced experimentation contrasting the performance of the proposed approach to other state-of-the-art solutions on an acoustic echo cancellation scenario confirms the effectiveness of our method for real-life applications.
\end{abstract}

\section{Introduction}
\label{Introduction}
Harmonic analysis has played a central role in modeling nonlinear physical operators in a mathematically tractable manner. There is a rich history in the literature of physical models using tools of harmonic analysis for acoustic applications like sound synthesis \cite{von1898vorlesungen, cremer1981physik, trautmann2012digital} or modeling different stages in electroacoustic systems to provide high quality audio signals in the real-life applications  \cite{muller1999digitale}.
In an electroacoustic system, an audio signal goes into multiple nonlinear distortions throughout its path in the playback and recording devices. 
These nonlinearities originate from the digital to analog converters, the magnetic hysteresis of an electrodynamic loudspeaker, power supplies, low quality microphones, and analog to digital converters \cite{dobrucki2011nonlinear, ravaud2010ranking}. Understanding these nonlinearities along the signal path is a valuable tool for diagnostics or signal quality improvement.

For dynamic nonlinear system identification, Volterra series based models \cite{cheng2017volterra}, Wiener-Hammerstein models \cite{wills2013identification} and Gaussian Process models \cite{principe2011kernel} have been used extensively. Another established method is nonlinear autoregressive exogenous model (ARX) \cite{zhang1997using, billings2005new}. More recently Bayesian neural networks have been used to directly identify nonlinear systems \cite{merkatas2021system}. In \cite{li2021fourier} it has been shown that deep neural networks (DNN) can successfully model complex physical systems. In this paper, we present an interpretable adaptive model for the \emph{online} identification of nonlinear systems. Our approach makes explicit use of harmonic analysis, a time-frequency representation of the signals of interest. The signals are divided into overlapping blocks. Then, the Fourier transform is applied on the windowed version of each block to extract the spectrum \cite{hlawatsch2013time}. The time-frequency representation allows to model non-stationary signals which are of interest in most practical applications such as in music, speech, and biomedical signals. Our model is stackable and physically interpretable. As in DNNs, in our model nonlinear layers are stacked and the network maps the input signal to the output through a composition of functions represented by these layers. We constrain the functions in every layer to be representable by a frequency band (subband) mixing matrix and delay operators. Every layer plays a distinct role in minimizing the cost function, a property that is ensured by decorrelating the input signals across the stages. This will be shown in section \ref{sect:lattice}.% The identification happens in a subband domain where we have a high time-frequency resolution. This allows to reveal the causality 

Interpretability in machine learning can help extract the relevant information from a machine learning model about the dependencies or what was learned by the model \cite{murdoch2019definitions}. Further, it helps justify the resulting model's complexity \cite{molnar2018interprtable}. %If we can find the minimum number of layers required to model a dynamic system, we can determine how complex the nonlinear model is. 
For example, our proposed structure is able to model a Doppler shift with a single layer given the explicit capability of the proposed structure to map across frequencies. Also validation of the physical plausibility of a machine learning model is enabled by an interpretable model and hence, the correctness and its behavior predictability \cite{samek2017explainable} can be analyzed. For example, if a system is a cascade of a strong nonlinear component, followed by a linear and finally another weaker nonlinear component, our proposed model will have this characteristic reflected in the pattern of the subband mixing matrix along the layers. The layers representing strong nonlinearity will be dense, the linear system will be identified as diagonal mixing matrix, and the weak nonlinearity will have a sparse mixing matrix.  Another benefit of interpretability is its use for generalization. Interpretability helps to identify incompleteness in input data or modeling limitations \cite{doshi2018considerations}. In fact, model interpretability makes it easier to reveal training data deficiency or to identify the model's limitations. This helps the generalization by exposing the system to a targeted training input or by systematically incorporating domain specific prior knowledge as regularization.

Finally, while %end-to-end learning in deep neural networks, most machine learning structures 
many black-box deep neural structures are very capable of modeling complex systems by finding the dependencies in the input/output data, they are limited in their ability to extract an explicit causal relationship between different components of the underlying system \cite{lecca2021machine}. This results in a slow adaptation capability when only a few components of the underlying system change. In some cases, they might even need a complete retraining. The structure we present in this paper explicitly reveals the causality between different components of the system under modeling. This helps fast track the dynamic system.

We call our proposed structure ``Neural Harmonium'' inspired by Smolensky's work which introduced the harmony principle as a foundation for a cognitive system \cite{smolensky1986information}. We are using the name because our system makes use of harmonic analysis as a tool to model the behavior of a dynamical system and therefore, simulate its ``Harmony''.
The proposed model describes the underlying system as a composition of independent cascaded nonlinear units operating on decorrelated input. As it will be shown, the benefit of decorrelating the input signal is the block-diagonalization of the Hessian. This enables efficient implementation of second order optimization techniques.

\section{Interpretable Nonlinearity Modeling}
Fourier transform expresses a time dependent signal by means of complex exponential functions. Since the eigenfunctions of any linear time-invariant (LTI) system are complex exponentials, any LTI has a diagonal representation in the frequency domain. In the present work, we make explicit use of the ability to model a nonlinear system using non-diagonal system representation in the frequency domain and show that this model can capture many practical nonlinearities. Let's say a set of $K$ frequency components given by $\mathbb{F}^i = \left \{ f^i_1, f^i_2, \dots, f^i_K\right \}$ constructs the excitation signal, $x(t)$, as
\begin{align}
	x(t) = \sum_{k=1}^K \check{X}_k(t) \sin(2\pi f_k^i t),
\end{align}
where $t$ is time, $\check{X}_k(t)$ is the piece-wise constant magnitude of the $k$-th frequency component.
% $w_m$ is the envelope of $m$-th frequency component. Let $h$ be the nonlinear system, 
In the presence of nonlinearity, the output of the system would have additional frequency components that did not exist in the excitation signal $x(t)$. The additional frequencies could be the harmonics (multiple integer of the excitation frequencies) or cross term components. For generality, let's say for a given excitation frequency set $\mathbb{F}^i$ the system produces a set of output frequencies $\mathbb{F}^o =  \left \{ f^o_1, f^o_2, \dots, f^o_{K_{\mathbb{F}^i}} \right \}$. The size of $\mathbb{F}^o$ is $K_{\mathbb{F}^i}$ which is a function of the set of excitation frequencies. The output, $y(t)$ can be written as 
\begin{align} \label{eq:cont-nl}
	g(x(t)) = \sum_{k = 1}^{K_{\mathbb{F}^i}} \check{Y}_k(t) \sin(2 \pi f^o_k t),
\end{align}
where $\check{Y}_k(t)$ is the piece-wise constant magnitude of the $k$-th frequency component. 
We assume the continuous-time excitation, $x(t)$ has a finite bandwidth and model the nonlinearity in discrete time-domain. Let $f_s$ denote the sampling frequency which is at least two times larger than the maximum frequency that exists in $x(t)$. Then the discrete time domain representation would be $x[n] = x(n / f_s)$ where $n$ denotes the discrete time index. We use a subband transform to divide a signal into a set of narrow-band signals. We can use various subband transformations to extract localized frequency content of the input. One may use Short-Time-Fourier-Transform (STFT) \cite{waldron2018introduction}, or a complex version of the  Modified-Discrete-Cosine-Transform (MDCT) \cite{CMDCT_1}, %Multi-Hop-Complex MDCT (MH-CMDCT) \cite{cmdct}, 
to name a few. However, it is important to have a filterbank which ensures the analyticity of the signal, as it is shown in section~\ref{sect:lattice}. %This can be achieved by applying a Hilbert transform on the STFT filterbank. %In the rest of the paper, we use STFT for its simplicity. 
Let $w_a[n]$ of size $N_w$ be an analysis window, then the analysis operation is given by 
\begin{align}
	X[k, l] = \sum_{n=lN_h}^{N_w + lN_h - 1} x[n] w_a[n - l N_h] e^{ - j (2 \pi \frac{k}{N_w}+\frac{\pi}{N_w})n},\nonumber
\end{align}
where $k$ is subband index, $l$ is the block time index, $N_h$ is the hop size. The synthesis operation is given by 
\begin{align}
	%	\text{\small$\displaystyle\tilde{x}[n] = \frac{1}{N_w} \sum_l w_s[n - lN_h]\sum_{k=0}^{N_w - 1} X[k, l] e^{  j 2 \pi k (n - l N_h) / N_w}$}\nonumber
	\text{\small$\displaystyle\tilde{x}[n] = \frac{1}{N_w} \sum_l w_s[n - lN_h]e^{  j  \frac{n\pi}{N_w}}\sum_{k=0}^{N_w - 1} X[k, l] e^{  j 2 \pi k \frac{(n - l N_h)}{N_w}}$},\nonumber
\end{align}
where $w_s[n]$ is the synthesis window pair of $w_a[n]$.
The subband representation allows us to analyze each bin in the output of a nonlinear system individually. Let's say $x[n]$ with subband representation $X[k, l]$ is the excitation to the nonlinear system and $y[n]$ with subband representation $Y[k, l]$ is the corresponding output of the nonlinear system. 
Since the time signal we are interested in is real and due to the analyticity property of the filterbank, we are interested only in the positive half of the signal spectrum.
With nonlinearity defined in (\ref{eq:cont-nl}), we define an activation set $\mathbb{A}_{k_o}$ to be a set of frequency bins such that if the excitation signal has non-zero values in this set the output would have non-zero value at bin $k_o$. 

Assuming the number of bins is $N_s$, then we have $N_s$ of such sets, $\mathbb{A}_0, \mathbb{A}_1, \dots, \mathbb{A}_{N_s-1}$. For the output bin $k_o$, the signals of the bins in $\mathbb{A}_{k_o}$ are combined using a multi-input-single-output (MISO) linear system. The MISO system may have memory of length $L$. This allows us to model the previous states of inputs that have an effect on the current output. Then, the output at bin $k_o$ can be written as a multi-channel convolution in the time-frequency domain over $L$ taps defined by
\begin{align} \label{eq:miso_conv}
	Y[k_o, l] = \sum_{k \in \mathbb{A}_{k_o}} \sum_{l^\prime = 0}^{L - 1} H_{k_o}[k, l^\prime] X[k, l - l^\prime],
\end{align}
where $H_{k_o}[k, l^\prime]$ is the $l^\prime$-th tap filter coefficient at the input bin $k$ for the output bin $k_o$. To estimate $H_{k_o}$ any multi-channel adaptive filter algorithm can be used. However, we propose a multi-channel Kalman based adaptive filter to estimate each of the linear filters, $H_{k_o}$ for $k_o=0, 1, \dots, N_s-1$. We use a neural network to identify the activation set, $\mathbb{A}_{k_o}$ for $k_o = 0, 1, \dots, N_s-1$.

\begin{figure*}[htp]
	\centering
	\includegraphics[scale=.24]{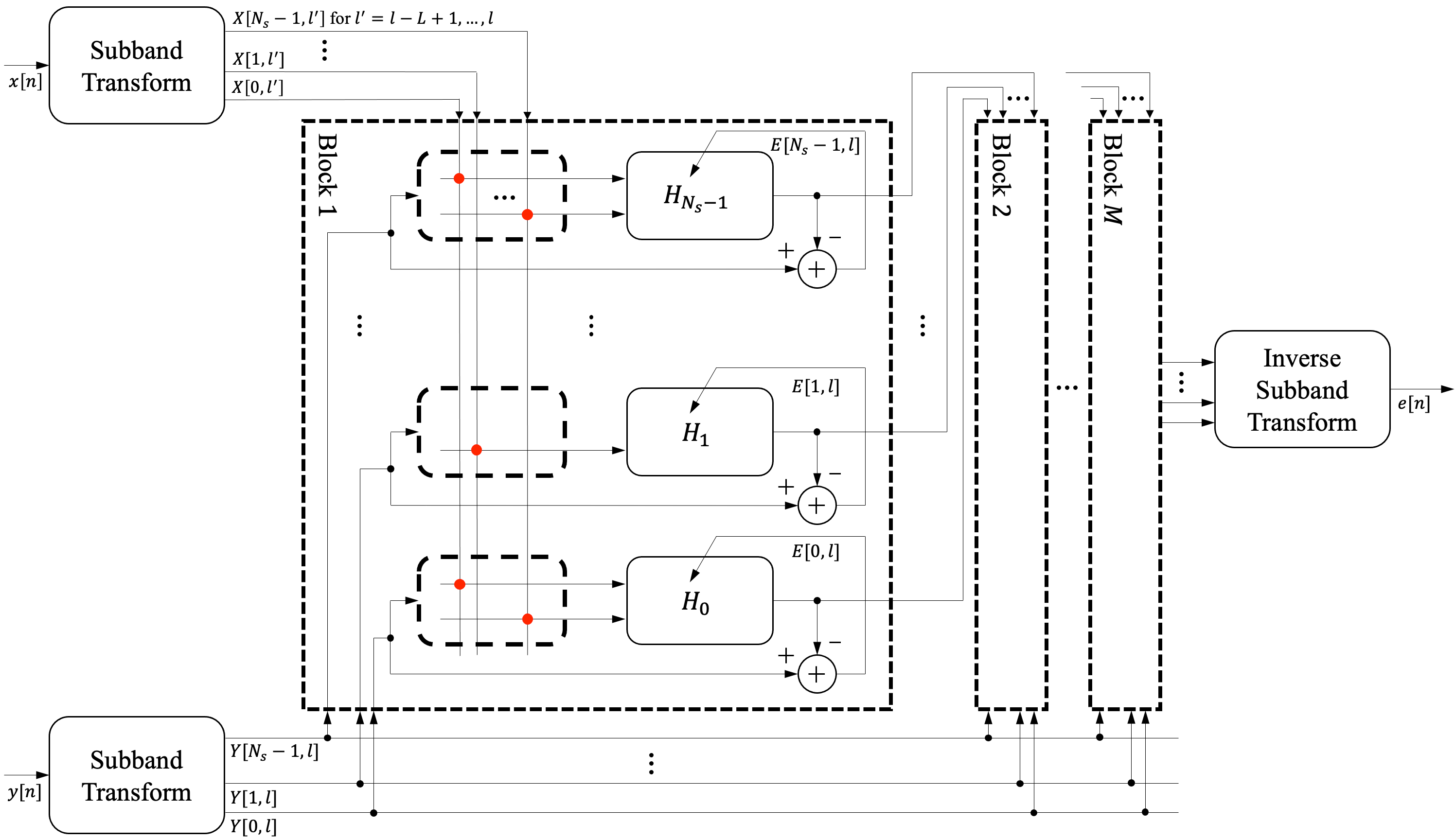}
	\vspace{0pt}
	\caption{This figure shows the proposed system for estimating nonlinearity. For each block index $l$, both the excitation and output signals are transformed into subband representations denoted by $X$'s and $Y$'s respectively. A neural network learns the connection between the output of the model and the input in subband domain. More specifically, the neural network decides where to put the red dot connections in this figure. After the connections are established, an adaptive filter tracks the changes in each individual subband by estimating the linear coefficients represented by $H$'s. At each subband, the difference of the actual nonlinearity and estimated nonlinearity is calculated and is used to updated the linear coefficient. This differences are denoted by $E$'s and are converted to the time domain error signal by using the corresponding inverse subband transform. The resulting time domain error signal is denoted by $e[n]$.}
	\label{fig:building_block}
\end{figure*}

\section{Single-stage Multichannel Dynamical System Modeling} \label{sec:ssmd}
In the following we consider multiple input single output (MISO) systems since any multiple input multiple output (MIMO) system can be decomposed into parallel MISO systems. Each MISO system in our model maps {{$N_s$}} subbands to an output subband say $k$ at the observation time $l$. 
\begin{align}
	\bm{h}_{k, l+1} = \bm{\theta}(\bm{h}_{k, l}, \bm{u}_{k, l}) + \bm{\nu}_{k, l},
\end{align}
where  $\bm{u}_{k, l}$ is the control input, $\bm{\nu}_{k, l}$ is the process noise with the covariance $\Gamma_{\Delta, k}^2:={\mathcal{E}}\{\bm{\nu}_{k, l} \bm{\nu}_{k, l}^H\}$ with $\mathcal{E}\{\cdot\}$ being the expectation and $\{\cdot\}^H$ denoting the Hermitian transpose. $\bm{h}_{k,l}=[{\bm{h}^{0}_{k,l}}^T,\ldots, {\bm{h}^{k_{N_s-1}}_{k,l}}^T]^T$ with $\{\cdot\}^T$ denoting the transpose operator. $\bm{h}^{k_i}_{k,l}=\left[H_{k,l}(k_i,0),\right.$$\left.\ldots, H_{k,l}(k_i,L-1)\right]^T$,  with $H_{k,l}(k_i,l^{\prime})$ denoting the snapshot of the filter at time $l$ and  $\bm{\theta}(\cdot,\cdot)$ describing the system dynamic. For the special case of linear Gauss-Markov model, the state transition equation reads
\begin{align}
	\bm{h}_{k, l+1} = \bm{A}_{k}\bm{h}_{k, l} + \bm{\nu}_{k, l}, \label{eqn:state} 
\end{align}
where $\bm{A}_{k}$ represents a state transition matrix for the output at subband $k$.
The measurement equation is given by
\begin{align}
	d_{k, l} = \bm{h}^{H}_{k, l} \bm{\mathcal{x}}_{l} + \varepsilon_{k, l}, \label{eqn:meas} 
\end{align}
where  $\varepsilon_{k, l}$ is the measurement noise for subband $k$ and at block $l$ with covariance $\xi^2_{k, l}:={\mathcal{E}}\{\varepsilon_{k, l}\varepsilon^*_{k, l}\}$ where $\{\cdot\}^*$ is the complex conjugate operator and $\bm{x}_{l}=[{\bm{x}_{0,l}}^T,\ldots, {\bm{x}_{K-1,l}}^T]^T$, $\bm{x}_{k,l}=\left[X[k,l],\right.$$\left.\ldots, X[k,l-L+1]\right]^T$. 

In general, obtaining the coefficients $\bm{h}_{k, l}$ involves solving a set of equations of the form
\begin{equation}
	\mathbf{Q}({\bm{h}}_{k, l})=\mathbf{0},
\end{equation} 
where $	\mathbf{Q}(\cdot)$ here is the gradient of the cost function $\mathcal{J}$
\begin{equation}
	\nabla_{{\bm{h}}} \mathcal{J}(\bm{h}_{k, l})=\frac{\partial \mathcal{J}(\bm{h}_{k, l})}{\partial \bm{h}_{k, l}}=\mathbf{0}.
\end{equation}
Using Taylor expansion around a certain point ${\bm{h}_{k,l}^{\ell-1}}$ up to the first order gives the approximation
\begin{align}
	\mathbf{Q}({\bm{h}}_{k, l})\approx&\left.\mathbf{Q}({\bm{h}_{k, l}})\right|_{{\bm{h}_{k, l}^{\ell-1}}} +\left[\frac{\partial}{\partial{\bm{h}_{k, l}}} \mathbf{Q}^{{T}}({\bm{h}_{k, l}})\right]_{{\bm{h}_{k, l}^{\ell-1}}}^{{T}}\nonumber\\
	&\qquad\cdot\left({\bm{h}_{k, l}} - {\bm{h}_{k, l}^{\ell-1}}\right).%=\mathbf{0},
\end{align}
Setting this approximation to zero leads to the Newton algorithm given by% $\mathbf{Q}({\bm{h}}_{k, l})=\nabla_{{\bm{h}}}\mathcal{J}\left(\bm{h}^{\ell-1}_{k, l}\right)$ and neglecting the residual term

\begin{equation}\hspace{-.15cm}
	\begin{array}{r}
		{\bm{h}_{k, l}^{\ell}}\hspace{-2pt}=\hspace{-1pt}{\bm{h}}^{\ell-1}_{k, l}\hspace{-2pt}-\hspace{-2pt}\left[\nabla_{{\bm{h}}} \hspace{-2pt}\nabla_{\bm{h}}^{{T}} \mathcal{J}\hspace{-2pt}\left(\bm{h}^{\ell-1}_{k, l}\right)\right]^{-1} 
		\nabla_{{\bm{h}}} \mathcal{J}\hspace{-2pt}\left(\bm{h}^{\ell-1}_{k, l}\right).
	\end{array}
\end{equation}
It is worth noting that recent research on second order optimization for deep neural networks uses the Fisher information matrix (FIM) as a corner stone for the optimization. FIM is related to the Hessian $\nabla_{{\bm{h}}} \nabla_{\bm{h}}^{{T}} \mathcal{J}$ in the sense that $\nabla_{{\bm{h}}} \nabla_{\bm{h}}^{{T}} \mathcal{J}$  is the expected Hessian of the loss function $\mathcal{J} $ under the training distribution while FIM is the expected Hessian of $\mathcal{J}$  under the model’s distribution  \cite{martens2015optimizing}. 
Exact second order methods require the inverse of the Hessian which can be computationally expensive depending on the size of the Hessian. In our proposed approach, we employ a neural architecture, which we will introduce in section \ref{sect:DNN}, to output the dependencies across subbands. This limits the size of the Hessian by modeling the significant cross dependencies.\\
In this study, we will focus on optimizing the system dynamic constrained cost function given by 
\begin{align} \label{eq:dd}
	\mathcal{J}(\bm{h}_{k, l+1}) &= \hat{\mathcal{E}} \left \{ || D_{k, l+1} - \bm{h}_{k, l+1}^H\bm{x}_{ l+1} ||_{\xi^2}^2\right. \nonumber\\
	& \left.+ \lambda || {\bm{h}}_{k, l+1} -  \bm{A}_k{\bm{h}}_{k, l} ||_{\Gamma_{\Delta}^2}^2 \right\},%\bm{\Upsilon}_k
\end{align}
%where $\bm{\Upsilon}_k = \left [ \bm{h}^T_{k, 0}, \bm{h}^T_{k, 1}, \dots  \right ]^T$ is a state vector with the history of $\bm{h}$, 
where $D_{k, l}$ is the desired output and subband $k$, and $\lambda$ is a Lagrangian.
It can be shown that the above introduced cost function can be solved using a Kalman filter \cite{Humpherys, Buchner}. Typically,  direct implementation of the Kalman estimation suffers from instabilities and is sensitive to ill-conditioning. Better numerical properties are obtained by expressing the Kalman filter in the covariance form, \cite{Haykin:2002}. The update equations can then be summarized as shown in Algorithm~\ref{alg:kalman} where $\bm{\mho}^2_{k,l}$ denotes the inverse of the Hessian.
\vspace{-.4cm}
\begin{algorithm}[htb]
	\caption{Multichannel Kalman Filter}\label{alg:kalman}
	{\bfseries Input:} $\bm{x}_{l}$, ${D}_{k,l}$ \\
	\textbf{Output:} $\hat{\boldsymbol{h}}_{k,l+1}$
	\begin{algorithmic}
		\STATE{$\eta_{e, k,l}^2 =\xi_{l}^2+\bm{x}_{l}^{H} \bm{\mho}_{k,l-1}^2 \bm{x}_{l}$,}
		\STATE{$\bm{k}_{k,l} =\eta_{e,k,l}^{-2} \bm{A}_{k,l} \bm{\mho}_{k,l-1}^2 \bm{x}_{l}$,}
		\STATE{${E}_{k,l} ={D}_{k,l}-\hat{\boldsymbol{h}}_{k,l}^{H}\bm{x}_{l}$,}
		\STATE{$\hat{\boldsymbol{h}}_{k,l+1} =\bm{A}_{k,l} \hat{\boldsymbol{h}}_{k,l}+\bm{k}_{k,l} {E}_{k,l}^*$,}
		\STATE{$\bm{\mho}_{k,l}^2 =\bm{A}_{k,l} \bm{\mho}_{k,l-1}^2 \bm{A}_{k,l}^{H}+   \Gamma_{\Delta,k}^2 -\eta_{e,k, l}^2\bm{k}_{k,l}\bm{k}_{k,l}^{H}$.}
		%\vspace{0pt}
	\end{algorithmic}
\end{algorithm}	
\vspace{-1cm}
\section{Multi-stage Dynamical System Modeling}
\subsection{General Optimization Criterion}
%Finding a control sequence to a multistage nonlinear dynamic system has been studied extensively in the literature especially with the success of deep neural networks. 
In the following, we review a general form for optimizing a control sequence for a multistage system, where the input of every stage is the output of the previous stage
\begin{align}\label{eq.multistage}
	\bm{y}(i+1)=&g^{i}(\bm{y}(i), \bm{H}(i)) ,\nonumber\\
	\phi(\bm{y}(M))=&\bm{0},
\end{align} where $g^i(\cdot, \cdot)$ is a convolution of a multiband signal with the MIMO finite impulse response (FIR) filter $\bm{H}$ in the subband domain as shown in Eq.~(\ref{eq:miso_conv}), $i$ denotes the stage index $i\in\{0,\ldots,M-1\}$, 
$\bm{y}(M)$ is the final output, and $\bm{y}(0)=\bm{x}$. The function $\phi(\cdot)$ is a loss function constraining the output of the final stage which can be seen as generalization of the first term in Eq.~(\ref{eq:dd}).
By employing Eq.~(\ref{eq.multistage}), a general cost function constraining the final output of the system and the control coefficients sequence $\bm{H}$  can be defined as
\begin{align}
	{\mathcal{J}}(\bm{H})&=\phi(\bm{y}(M))+\sum_{i=0}^{M-1}\left[\mathcal{L}^{i}[\bm{y}(i), \bm{H}(i)] + \lambda^{T}(i+1)\right.\nonumber\\
	&\left. \cdot \left\{g^{i}(\bm{y}(i),\bm{H}(i))-\bm{y}(i+1)\right\}\right].
\end{align}
This cost function can be seen as a generalization of the cost function in Eq.~(\ref{eq:dd}) to the multi-stage case,
where $\mathcal{L}^i[\cdot,\cdot]$ represents a constraint on the model parameters at stage $i$ given an input at a stage $i$, which can be understood a generalization of constraints that can be put in place on the system or signal such as the state transition constraint presented in the second term of Eq.(\ref{eq:dd})and $\lambda$ is a Lagrangian.
If  the sequence $\lambda(i)$ is chosen such that \cite{bryson2018applied}
%	\lambda^{T}(i)-\frac{\partial J^{\prime^{i}}}{\partial \bm{y}(i)}=&0 \Rightarrow 
\begin{align}\label{eq:lambda}
	\lambda^{T}(i)=\frac{\partial \mathcal{L}^{i}}{\partial \bm{y}(i)}+\lambda^{T}(i+1) \frac{\partial g^{i}}{\partial \bm{y}(i)},
\end{align}
%for $ i=0, \ldots, M-1$, and 
with the boundary condition
\begin{equation}
	\lambda^{T}(M)=\frac{\partial \phi}{\partial \bm{y}(M)}.%+\mu\frac{\partial \psi}{\partial \bm{y}(M)}
\end{equation}
It can be shown that a minimum of the cost function can be found if the following condition is satisfied \cite{bryson2018applied}
%\begin{equation}
%	d \bar{J}=\sum_{i=0}^{N-1} \frac{\partial J^{\prime^{i}}}{\partial h(i)} d h(i)+\lambda^{\tau}(0) d x(0)
%\end{equation}
%
%\begin{equation}
%\frac{\partial J^{\prime^{i}}}{\partial h(i)}=0
%\end{equation}
\begin{equation}\label{eq:cond}
	\frac{\partial \mathcal{L}^{i}}{\partial \bm{H}(i)}+\lambda^{T}(i+1) \frac{\partial g^{i}}{\partial \bm{H}(i)}=0. %\quad i=0, \ldots, M-1.
\end{equation}
%Hence,
%\begin{equation}
%	\lambda^{T}(i+1)=\left[\lambda^{T}(i)-\frac{\partial L^{i}}{\partial %\bm{y}(i)}\right]\left[\frac{\partial f^{i}}{\partial \bm{y}(i)}\right]^{-1}.
%\end{equation}
An iterative approach to fulfill the condition in Eq.~(\ref{eq:cond}) is the Newton method which requires the estimation of the Hessian matrix. From Eq.~(\ref{eq:lambda}) and (\ref{eq:cond}), it can be seen that the independence of the inputs across the different stages will result in a block-diagonal Hessian. In the literature, multiple approaches can be found to efficiently approximate the Hessian matrix, e.g., \cite{martens2015optimizing}, \cite{ollivier2015riemannian}. In \cite{le2007topmoumoute} the authors suggest approximating it as block-diagonal matrix and ignoring terms across different layers. Here, we use the idea that is already mentioned in section \ref{sec:ssmd} to limit the size of the Hessian at every stage using a dedicated neural network. Further, we decompose our system in a manner that makes cross terms in the exact Hessian across layers vanish naturally. The idea is to constrain the input of each stage to be orthogonal to the output of subsequent stages. Hence, the Hessian will be naturally block-diagonal. This can be achieved by using the so-called lattice structure \cite{dsp}. 

\subsection{Decomposition of Multistage Dynamical System}\label{sect:lattice}
\begin{figure}[htp]
	\centering
	\includegraphics[scale=.185]{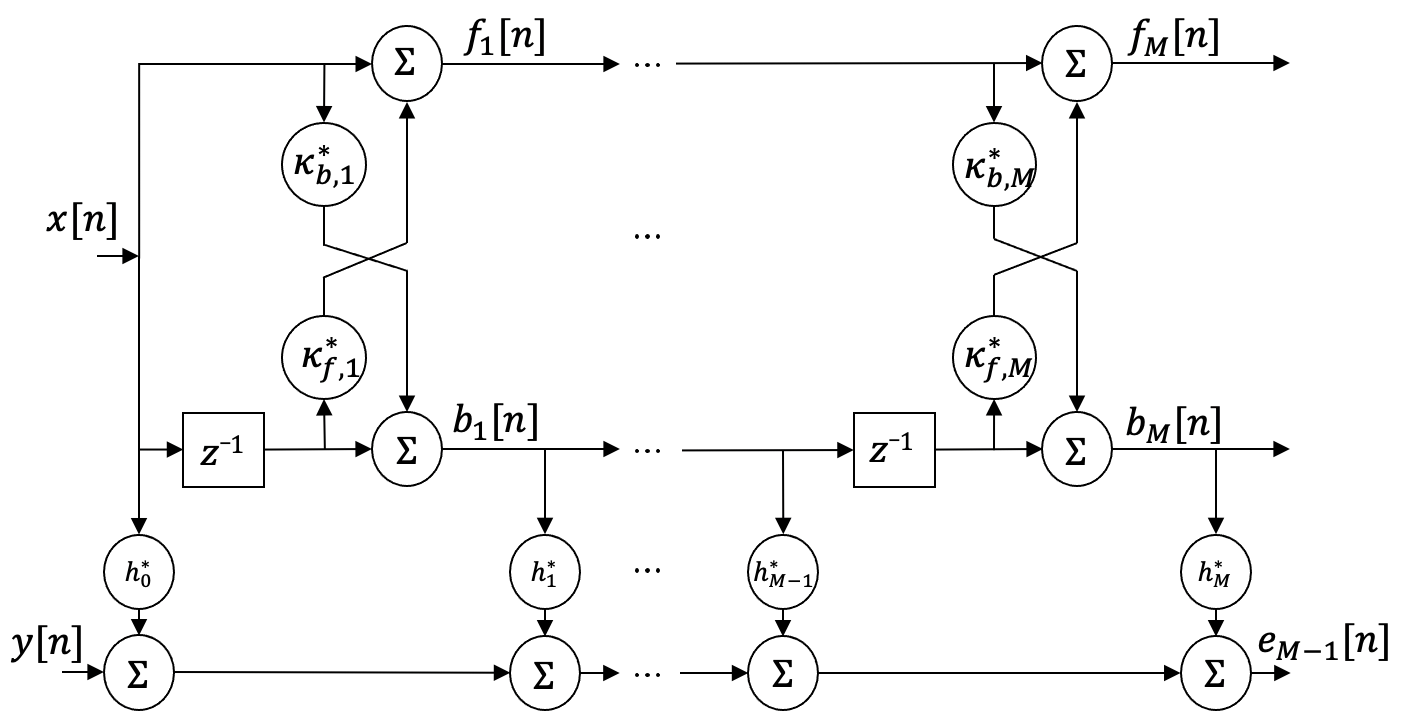}
	\vspace{-8pt}
	\caption{This figure shows the architecture of an adaptive filter estimating a convolutive mixing system in lattice form.}
	\label{fig:lattice}
\end{figure}

The proposed model for modeling nonlinearity is a multichannel convolutive mixing system where the individual channels are subband representations of the signals. The operation of convolution can be decomposed into a cascade of independent stages in a lattice structure. Employing lattice structure for single channel adaptive filtering tasks leads to efficient  implementations \cite{Cioffi, Haykin:2002}. The main idea of adaptive lattice filters is to transform the input signal into uncorrelated backward and forward prediction errors, $\bm{b}_m, \bm{f}_m$ respectively and estimating a single stage filter which minimizes the error between the target signal and its output in the least squares sense. The transformation into prediction errors ensures the orthogonality between the inputs of consecutive stages and hence, the independence of the inputs across the different stages. Figure~\ref{fig:lattice} depicts the structure of the model we use for the adaptive filter.  
%\subsection{Lattice structure decomposition}
It can be seen from Figure~\ref{fig:lattice} that in the $z$-domain the forward and backward prediction errors are coupled with $\bm{\kappa_{b,n}}$ and $\bm{\kappa_{f,n}}$ as
\begin{equation}\label{eq_lattice}
	\left[\begin{array}{l}
		\bm{b}_{m+1}(z) \\
		\bm{f}_{m+1}(z)
	\end{array}\right]=\left[\begin{array}{cc}
		z^{-1} \bm{I} & \bm{\kappa}_{b,m}^\ast \\
		z^{-1} \bm{\kappa}_{f,m}^\ast & \bm{I}
	\end{array}\right]\left[\begin{array}{l}
		\bm{b}_{m}(z) \\
		\bm{f}_{m}(z)
	\end{array}\right],
\end{equation}
where $\bm{I}$ is the unity matrix and $z^{-1}$ refers to a unit delay operator. $\bm{\kappa}_{b,m}$ and $\bm{\kappa}_{f,m}$ are calculated as solutions to a single stage forward/backward prediction problem using the algorithm presented in Algorithm~\ref{alg:kalman} multiple times. The details of the algorithm are given in Appendix \ref{appendix:lattice_kalman}. Hence, we decompose the multistage system into cascaded single stages systems with $L=1$.

Introducing the quantity $\mathbf{R}_{m}$ as the least squares solution to the equation
\begin{equation}\label{eq:R}
	\bm{f}_{m}=z^{-1}\mathbf{R}_{m}\bm{b}_{m},
\end{equation}
which can be explained as prediction errors transfer function. Substituting it in Eq~(\ref{eq_lattice}) leads to the recursion
\begin{equation}\label{eq:schurr}
	\mathbf{R}_{m+1}(z)=\left(\mathbf{R}_{m}(z)+\bm{\kappa}^\ast_{f,m}\right)\left(\bm{I}+\bm{\kappa}^\ast_{b,m} \mathbf{R}_{m}(z)\right)^{-1}.
\end{equation}
The recursion in Eq.~(\ref{eq:schurr}) is a multidimensional version of the Schurr recursion which is often encountered in algorithms for inverse scattering or layered media modeling \cite{bruckstein1987inverse}. This recursion formula suggests modeling $\mathbf{R}$ as a composition of conformal maps \cite{winebrenner1998linear}. This is significant as it guarantees the causality, stability, and uniqueness of the system model given the initial forward and backward predictions (at the first stage) are analytical.  
%An important special case of  multistage system identification is modeling layered media using  time-dependent inverse scattering. Algorithms for layered media modeling makes strong use of causality.
When a signal is given as a function of frequency, causality is reflected in the analyticity of the data in the complex upper half-plane. In our approach, we ensure this requirement by applying the Hilbert transform to the original filterbank. That is, we enforce the imaginary part of the spectrum to be the negative Hilbert transform of the real part of the spectrum.
%%In the frequency domain, splitting into the past and future is provided by the linear Riesz transform

\subsection{Model to Estimate the Sparsity Structure of the Inverse Fisher Information Matrix}\label{sect:DNN}

\subsubsection{Neural Architecture}
\begin{figure*}[htp]
	\centering
	\includegraphics[scale=.28]{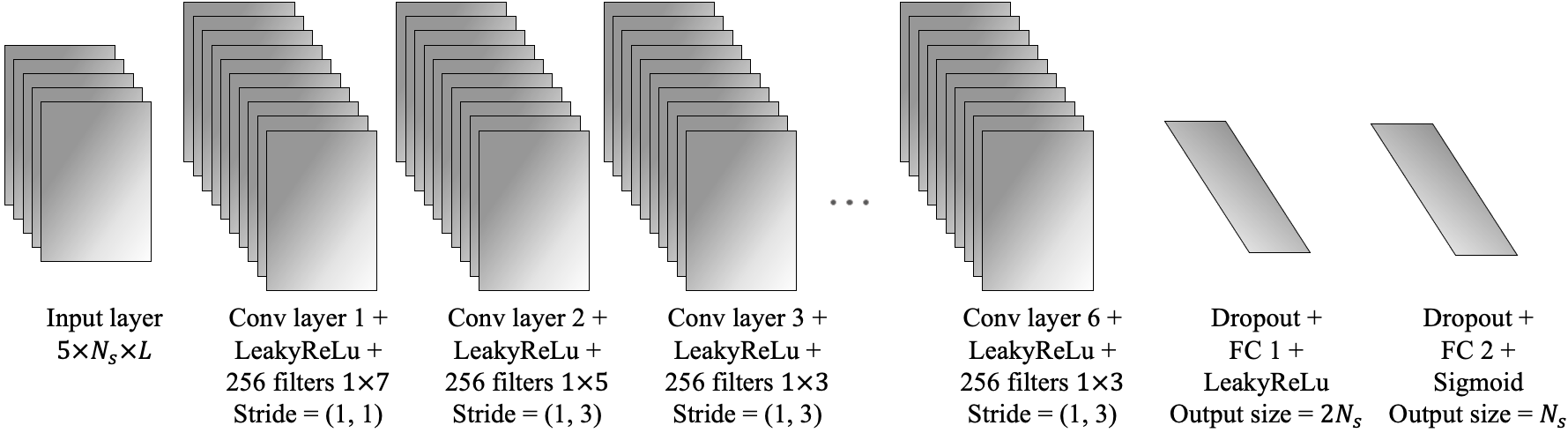}
	\vspace{-4pt}
	\caption{The architecture of the proposed neural network for dependency detection.}
	\label{fig:dnn_arch}
\end{figure*}

We use a deep neural network to identify the dependencies within the subbands. The input to the network comes from the last $L$ subband representations of the excitation signal, $X[\cdot,~ l^\prime]$ for $l^\prime = l-L+1, \dots, l$ and the last measurement signal's subband representation, $Y[\cdot,~ l]$ repeated $L$ times. This adds a memory to the network and lets the network identify the long-term dependencies. We treat the real and imaginary parts as separate network inputs. As an additional input, we use the absolute value of the cross-correlation between the subband representations of the excitation and the measurment. Since the subband representation has $N_s$ bins and we use $L$ such representations, the dimensionality of the input would be $3$ of form $5 \times N_s \times L$. The role of neural network is to decide which bins in the excitation signal contribute to which bins of the measurement. That is, for each nonlinearity's output bin $k^\prime$ the neural network finds a set like $\mathbb{K}_{k^\prime} = \left \{ k_0, k_1, \dots, k_{N_{k^\prime}} \right \}$ that contributes to $Y[k^\prime, \cdot]$ where $N_{k^\prime}$ is the total number of bins contributing to the output signal at bin $k^\prime$.
In practice for the output bin $k^\prime$, the network predicts a vector of $0$'s and $1$'s of size $N_s$ where $1$ at $k$-th place indicates the output bin $k^\prime$ has dependency to the excitation bin $k$ and $0$ shows the the nonlinearity output bin $k^\prime$ is independent of the excitation bin $k$. The same network is used to decide the dependency for all output subbands and similarly, for all stages. %The details of the network architecture are depicted in Appendix \ref{appendix:arch}. 
The architecture of the proposed neural network for dependency detection is shown in Figure \ref{fig:dnn_arch}. The input layer has $5$ channels which are obtained from the excitation signal and the output of the nonlinear system. Two channels are for the real and imaginary parts of the subband representation of the excitation, two channels are for the real and imaginary parts of the subband representation of the nonlinear system output, and one channel is for the absolute value of the cross correlation of the two subband representations. We use the last $L$ subband representations of the excitation signal $X$ but only the last subband representation of the nonlinearity output $Y$. Therefore before feeding $Y$, we repeat it $L$ times to match the dimension of $X$ and $Y$. This way we feed the input of size $5 \times N_s \times L$. Since the neural network is finding the relation within subbands, we only convolve the data in the time direction and not in the subband direction. The first convolutional layer (Conv layer 1) has strides of $(1, 1)$ but the remaining convolutional layers have strides of $(1, 3)$. The kernel size of the first convolutional layer is $(1, 7)$. The second convolutional layer has a kernel size of $(1, 5)$, and the remaining layers have kernel size of $(1, 3)$. We use four such layers. There are two fully connected layers with output sizes of $2N_s$ and $N_s$. All the layers use Leaky Rectified Linear Unit (LeakyReLu) at the output except the last layer which uses sigmoid to limit the output within $[0, 1]$. During the training, with each fully connected layer, we use a dropout layer for regularization. We set the probability of an element to be zeroed to $0.1$ for the dropout layers.
\subsubsection{Training}
The training dataset is generated using the following procedure. For each subband bin $k=0,\dots, N_s -1$, a random signal is generated and filtered by a random filter of random length to form the excitation input. Note that we generate the excitation signal in the subband domain not in the time domain. Then, a sparse dependency map of $1$'s and $0$'s is generated\footnote{The probability of a value in the map is $1$ is set to $0.3$ in our design}. Only if the value of the dependency map is $1$, then the excitation is convolved with another random channel and the result is added to the output. To discard the initialization values of the buffers in the filters, the last $L$ samples from all of the subband signals are used for training.  Finally, a simulated typical measurement noise is added to the output. The excitation and nonlinearity output forms the input to the neural network and the dependency mapping is used as label during the training of the neural network. Note that the task of the neural network is to detect if there is a dependency between an excitation bin and the output bin. In other words, it is solving a detection problem. Therefore, we train the parameters of the neural network by minimizing the binary cross entropy of the network prediction and the actual mapping.

\begin{figure}[htp]
	\centering
	\includegraphics[width=0.45\textwidth]{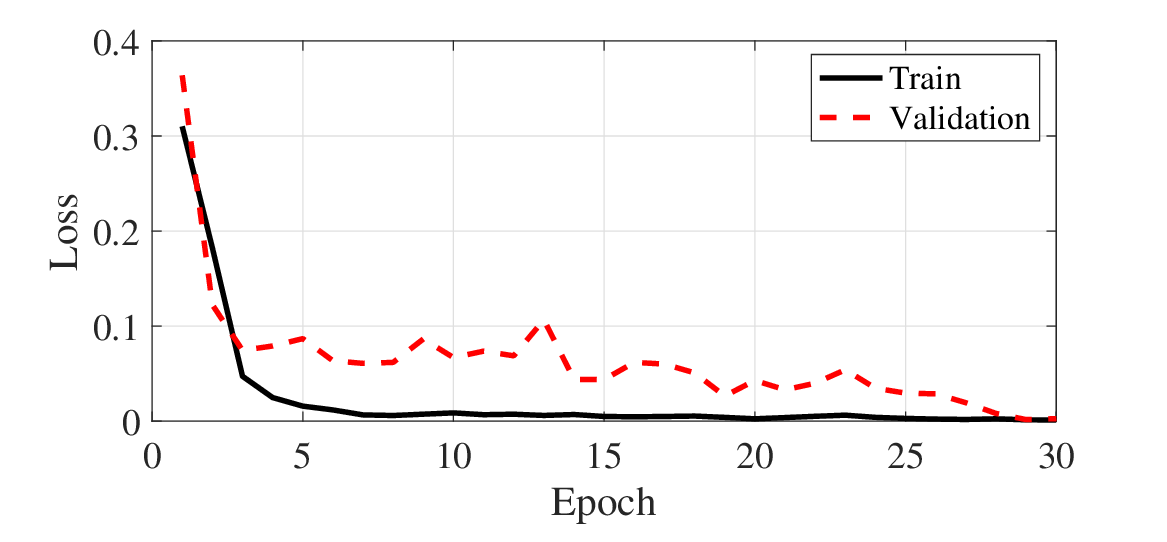}
	\vspace{-8pt}
	\caption{An example of training and validation loss for $N_w = 32$.}
	\label{fig:training_epochs}
\end{figure}

For the optimization algorithm, we use Adam, \cite{kingma2014adam}, with a learning rate of $1\mathrm{e}{-5}$, where $\beta_1 = 0.9$ and $\beta_2 = 0.999$. 
An example of training loss is shown in Figure \ref{fig:training_epochs}. An independent validation set has been created with $25\%$ the size of the training set.
\section{Results}
\label{Results}
In this section, we show the efficiency of Neural Harmonium in identifying different nonlinearities systems through a set of experiments. Additional illustrating experiment can be found in the appendix C.
\subsection{Magnetic Hysteresis as Nonlinearity}
In this experiment, we use the hysteresis model described in Appendix \ref{app:B} of hysteresis to simulate the nonlinear effect of a loudspeaker system. Then the loudspeaker signal excites a linear acoustic channel with reverberation time (RT60) of $200$ (ms) and captured by an ideal microphone.
The whole system is excited by a speech signal. We compare the modeling error of our method to Wiener-Hammerstein baseline model, with a linear part of the same order as our proposed model and input and output polynomial nonlinearities of order 10 \cite{wills2013identification}, a Gaussian process regression (GPR) using kernel RLS algorithm and Mat\'{e}rn kernel 32 \cite{principe2011kernel}, and a nonlinear autoregressive exogenous model (ARX) with a wavelet network as  \cite{zhang1997using, billings2005new}
The Wiener-Hammerstein model output achieved modeling error of  $-4.75$ dB, the GPR $-3.45$ dB, and the ARX $-2.99$ dB, whereas our proposed model outperformed all three models significantly by achieving $-15.93$ dB.

\begin{figure}[htp]
	%	\vspace{-12pt}
	\centering
	\includegraphics[width=0.5\textwidth]{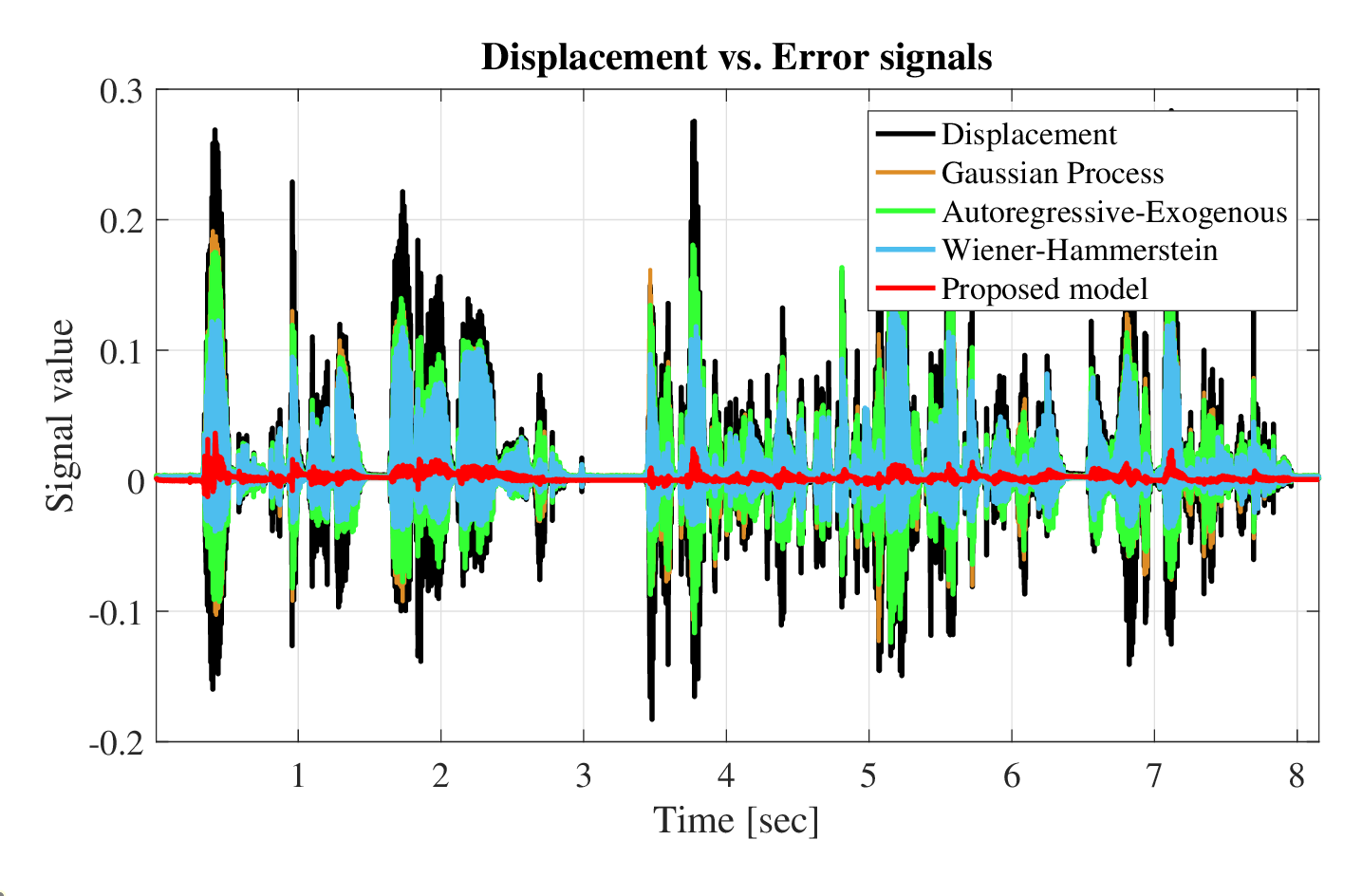}
	\vspace{-28pt}
	\caption{The output of the hysteresis model to a speech excitation signal. The modeling error for this system is $-15.93$ dB.}
	\label{fig:modeling_error_hys}
\end{figure}

% and pass it through a measured room impulse response with $T60$ of 400 millisecond.
The system creates a displacement signal which is shown in black in Figure \ref{fig:modeling_error_hys}. We use the excitation signal and the displacement signal to estimate the nonlinearity. For Neural Harmonium the total number of bins $N_s$ is set to $32$, hop size $N_h$ to $16$, and number of stages $M$ to $15$. We define the \emph{modeling error} in decibel (dB), as
%\begin{align} \label{eq:delta}
$\delta = 10 \log_{10} ( {\sum_n \left | e[n] \right |^2} / {\sum_n \left | y[n] \right |^2}  )~~\text{[dB]} $.
%\end{align}
The modeling error $\delta$ is $-15.93$ dB when we use Neural Harmonium to model this system. The resulting error signal is shown in red in Figure \ref{fig:modeling_error_hys}. The result shows the efficiency of the proposed method in modeling this nonlinearity.

\begin{figure}[htp]
	\centering
	\includegraphics[width=0.4\textwidth]{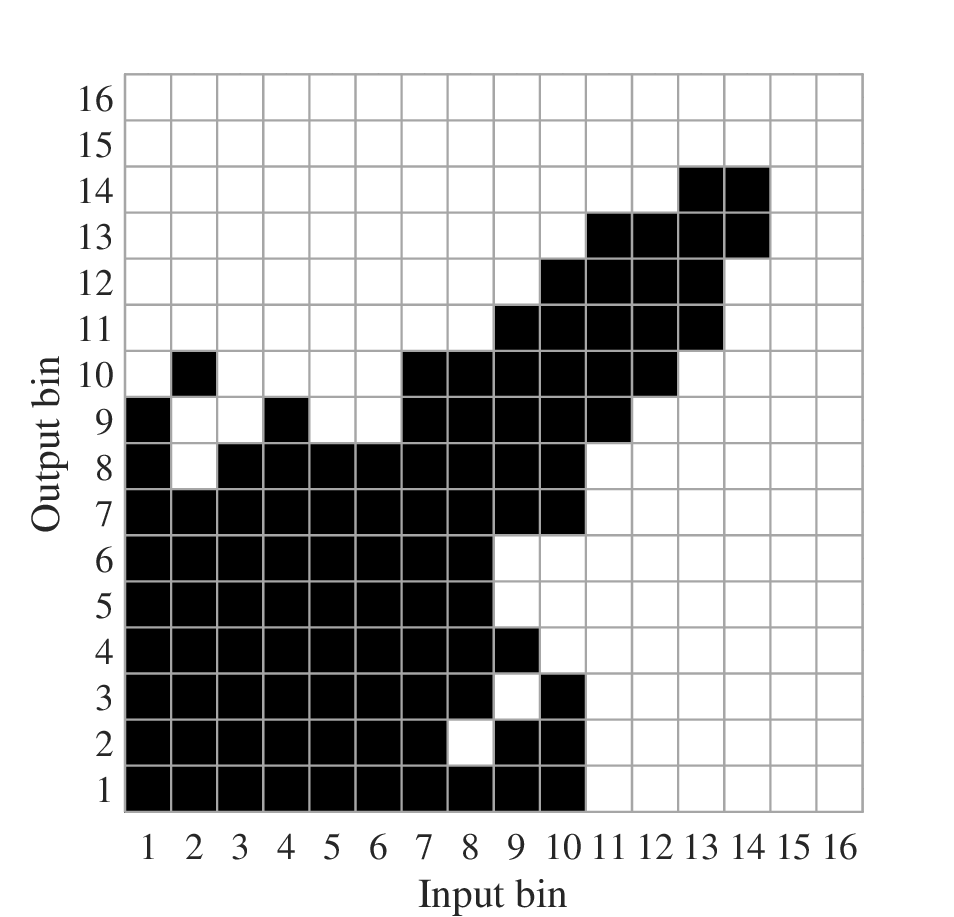}
	\vspace{-10pt}
	\caption{The dependency map of the hysteresis model.}
	\label{fig:hys_dependency_map}
\end{figure}
In Figure~\ref{fig:hys_dependency_map}, the average of the dependency map over all of the stages is depicted. It shows how the interdependence of the low frequency components with higher bands weakens as the frequency increases. The higher the frequency, the less nonlinear the system becomes. In this experiment, the limited bandwidth of the excitation signal resulted in the last two bins not being activated. 
Frequency modulators are other examples of nonlinear systems which can greatly illustrate the ability of our approach to interpret the underlying system as shown in Appendix \label{app:modulation}.

\subsection{Acoustic Echo Cancellation}
In teleconferencing applications Echo Control (EC) is required to remove echoes captured by the microphone while keeping the near-end signal intact \cite{benesty2013advances}. One of the challenges with Acoustic Echo Cancelers (AEC) is that the audio reproduction and capturing hardware are highly nonlinear. The nonlinearity comes from the digital-to-analog converter, the power amplifier, and the loudspeaker's magnet hysteresis in the reproduction path and comes from the microphone and the analog-to-digital converter in the capturing path. 
With the degree of nonlinearity in these components, any linear processing is insufficient to completely cancel the echo. Here we show the efficiency of the Neural Harmonium as a nonlinear AEC. We use the year $2022$ acoustic echo cancellation challenge dataset shared at \cite{microsoft_AEC_challenge}. We compare the performance of Neural Harmonium with the performance of WebRTC's open source EC which is used in many teleconferencing applications \cite{WebRTC_AEC}. WebRTC's EC has a dedicated component to combat nonlinearities in the system. Furthermore, we include in the comparison a more recent echo control system based on a perceptually motivated interference suppression architecture \cite{valin21}. The same architecture has won the first place in the 2021 echo cancellation challenge \cite{icassp21_challenge}. In order to be able to compare both approaches, the network in \cite{valin21} has been implemented without the pitch filtering subsystems as we are not focusing on extracting only a speech signal but rather suppress the echo and keep the near-end unaltered. Similar to the training procedure suggested in  \cite{valin21}, the network was trained on 120 hours of clean speech data along with 80 hours of various noise types and the echo was generated using both synthetic and real room impulse responses employing a typical audio processing application motivated augmentation process.
We use echo return loss enhancement ($\ERLE$) during the far-end single talk which is the negative of model error for evaluating echo suppression. That is $\ERLE = -\delta$.
We experimented with speech near-end signal as well as music. For the speech signal case, we use the perceptual evaluation of speech quality ($\PESQ$) score between the error signal and the actual near-end signal \cite{pesq}\footnote{If the actual $\PESQ$ is $x$, then the normalized $\PESQ$ would be $(x + 0.5) / 5$.}. For Neural Harmonium, we set the window size, $N_w$ to $256$, the hop size $N_h$ to $128$, and  $M$ to $60$. The average $\ERLE$ and $\PESQ$ is shown in Table \ref{table:aec_performance1}. 
%The result of using these two methods on $100$ samples are shown in Figure \ref{fig:aec_performance}. 
Neural Harmonium outperforms WebRTC in both $\ERLE$ and $\PESQ$ and performs objectively on par with the DNN based echo suppression network. We further use the subjective measure of mean opinion score (MOS) to evaluate the resulting near-end signal quality \cite{mos1996, mos1998}. The results for MOS are obtained from $400$ speech near-end real recordings shown in right column of Table \ref{table:aec_performance1}. For the music near-end signal, again 400 music samples have been mixed as near-end signal and the results are shown in Table \ref{table:aec_performance2}. Each output test sample is $10$ second long and rated by $10$ listeners from the pool of the experiment participants. The Neural Harmonium receives a higher score compared to the WebRTC’s EC but has been ranked subjectively $2^\mathrm{nd}$ after the pure DNN architecture for the speech near-end signal, however, for music near-end, Neural Harmonium has been ranked subjectively $1^\mathrm{st}$. Hence, the proposed approach doesn't compromise quality at the cost of interpretability.% \vspace{-10pt}

\begin{table}[!h]
\caption{Comparing Neural Harmonium performance to other state-of-the-art DNN and classical DSP based systems in canceling echo during speech double-talk scenario using objective and subjective metrics. $\ERLE$ in dB and normalized $\PESQ$ are used for the objective evaluation. MOS indicating the subjective evaluation.}
\label{table:aec_performance1}
\centering
%	\vspace{-4pt}
{
	\begin{tabular}{ c | c  | c | c}
		\toprule
		&  $\ERLE$ &  $\PESQ$ & MOS\\
		\midrule
		Neural Harmonium & $35.79$ & $0.43$ & $3.24$ \\
		\midrule
		WebRTC & $20.57$ & $0.33$ & $2.81$\\
		\midrule
		\parbox[t]{3cm}{Interference Suppression based on \cite{valin21}}& $35.95$ & $0.44$ & $3.33$ \\
		\bottomrule
\end{tabular}}
\end{table}
\begin{table}[!h]
\caption{Neural Harmonium performance contrasted to other state-of-the-art DNN- and classical-DSP-based systems in canceling echo during  double-talk scenario with music as near-end signal. The used subjective metric is MOS as obtained from a crowd-sourced listening experiment.}
\label{table:aec_performance2}
\centering
%	\vspace{-4pt}
{
	\begin{tabular}{ c | c }
		\toprule
		& MOS\\
		\midrule
		Neural Harmonium &  $2.78$ \\
		\midrule
		WebRTC &  $2.53$\\
		\midrule
		\parbox[t]{3.5cm}{Interference Suppression based on \cite{valin21}}& $2.71$ \\
		\bottomrule
\end{tabular}}
\end{table} 

\section{Conclusion}
\label{Conclusion}
In this paper, we presented a novel hybrid stackable structure which combines deep neural networks with elements of classical adaptive filtering. We use this structure to identify nonlinear systems online. The system identification employs harmonic analysis tools by performing the identification in a time-frequency framework. Experiments confirmed the capability of the proposed system in modeling nonlinear multistage dynamical systems for real-life applications.

\bibliography{nl_sysid}
\bibliographystyle{aaai23}

%%%%%%%%%%%%%%%%%%%%%%%%%%%%%%%%%%%%%%%%%%%%%%%%%%%%%%%%%%%%%%%%%%%%%%%%%%%%%%%
%%%%%%%%%%%%%%%%%%%%%%%%%%%%%%%%%%%%%%%%%%%%%%%%%%%%%%%%%%%%%%%%%%%%%%%%%%%%%%%
% APPENDIX
%%%%%%%%%%%%%%%%%%%%%%%%%%%%%%%%%%%%%%%%%%%%%%%%%%%%%%%%%%%%%%%%%%%%%%%%%%%%%%%
%%%%%%%%%%%%%%%%%%%%%%%%%%%%%%%%%%%%%%%%%%%%%%%%%%%%%%%%%%%%%%%%%%%%%%%%%%%%%%%
\clearpage\newpage
\appendix
%\onecolumn
%\setcounter{page}{1}

%\begin{center}
	\textbf{\Large{Appendices to Neural Harmonium}}
%\end{center}
\normalsize

\section{Adaptive Filtering in Lattice Structure} \label{appendix:lattice_kalman}
In this section, a multichannel Kalman adaptive filtering in the lattice form is explained. To calculate $\bm{\kappa}_b$, $\bm{x}_{l}$ is set to the forward prediction error and ${D}_{k,l}$ is set to the backward prediction error from the previous stage. 
To calculate $\bm{\kappa}_f$, $\bm{x}_{l}$ is set to the backward prediction error from the last stage and $\bm{d}_l=[{D}_{0,l},\ldots,{D}_{k,l},\ldots,D_{N_s-1,l}]^T$ is set to the forward prediction error from the last stage. 
Similarly, the coefficients $\hat{\bm{H}}$ at every stage are calculated by setting $\bm{x}_{l}$ to the backward prediction error and ${D}_{k,l}$ to the residual error between the actual desired system output and the previous stages' output. 

Throughout this algorithm, we assume that the dependency map (the output of the DNN in section~\ref{sect:DNN}) at every stage $m$ is the same between iteration $l$ and $l-1$. In the practical implementation, a change of the dependency map between two iterations is handled by a secondary system which allows dependency changes. During the periods in which the measurement noise is stationary, the respective stage is copied from the secondary system to the primary system once the secondary filter is providing an error lower than the primary filter. We further assume, the process noise covariance matrix $\Gamma_{\Delta,k}^2$ to be identical for all MISO subcomponents of the MIMO system and call it $\Gamma_{\Delta}^2$. We also call the MIMO error at stage $m$, $\bm{e}_{m,l}$.    
\begin{algorithm}[htb]
\caption{Multichannel Kalman Filter in Lattice Form}\label{alg:lat_kalman}
\text{Initialization}
\begin{align}
	\begin{aligned}
		\bm{f}_{0,l}&=\bm{b}_{0,l}=\bm{x}_l,\\
		\bm{e}_{0,l}&=\bm{d}_{l},\\
		\bm{\mho}_{f,m,0}^2 &=\bm{\mho}_{f,m,0}^2 =\bm{I},\\
		\hat{\boldsymbol{\kappa}}_{b,m,0}=\hat{\boldsymbol{\kappa}}_{f,m,0}&=\hat{\boldsymbol{H}}_{m,0}=\bm{k}_{f,m,0}=\bm{k}_{b,m,0}=\bm{b}_{m,-1}=\bm{0},\\ &\forall m\in\{0,\ldots,M-1\}.\nonumber
	\end{aligned}
\end{align}
\text{Estimation of measurement noise covariance}
\begin{align}
	\begin{aligned}
		\xi_{l}^2=\frac{1}{M}\sum^{M-1}_{m=0}|\bm{e}_{m,l}-\hat{\bm{H}}_{m,l}^H\bm{b}_{m,l}|^2,\nonumber
	\end{aligned}
\end{align}
\text{Forward prediction update at stage $m$}
\begin{align}
	\begin{aligned}
		\bm{f}_{m+1,l} &=\bm{f}_{m,l}-\hat{\boldsymbol{\kappa}}_{f,m,l}^{H}\bm{b}_{m,l-1}, \\
		\hat{\boldsymbol{\kappa}}_{f,m,l+1} &=\bm{A}_{k,l} \hat{\boldsymbol{\kappa}}_{f,m,l}+\bm{k}_{b,l} \bm{f}_{l}^H, \nonumber
	\end{aligned}
\end{align}
\text{Backward prediction update at stage $m$}
\begin{align}
	\begin{aligned}
		\eta_{f,e, m,l}^2 &=\xi_{l}^2+\bm{f}_{l}^{H} \bm{\mho}_{f,m,l-1}^2 \bm{f}_{m,l}, \\
		\bm{k}_{f,m,l} &=\eta_{f,e,m,l}^{-2} \bm{A}_{l} \bm{\mho}_{f,m,l-1}^2 \bm{f}_{m,l},  \\
		\bm{b}_{m+1,l} &=\bm{b}_{m,l-1}-\hat{\boldsymbol{\kappa}}_{b,m,l}^{H}\bm{f}_{m,l}, \\
		\hat{\boldsymbol{\kappa}}_{b,m,l+1} &=\bm{A}_{l} \hat{\boldsymbol{\kappa}}_{b,m,l}+\bm{k}_{f,m,l} \bm{b}_{m+1,l}^H, \\
		\bm{\mho}_{f,m,l}^2 &=\bm{A}_{l} \bm{\mho}_{f,m,l-1}^2 \bm{A}_{l}^{H}+   \Gamma_{\Delta}^2 -\eta_{f,e,m,l}^2\bm{k}_{f,m,l}\bm{k}_{f,m,l}^{H}, \nonumber
	\end{aligned}
\end{align}
%\algstore{myalg}
%\end{algorithm}
%\begin{algorithm}[t]
%\algrestore{myalg}
\text{Joint-process estimation at stage $m$}
\begin{align}
	\begin{aligned}
		\eta_{b,e, m,l}^2 &=\xi_{l}^2+\bm{b}_{l}^{H} \bm{\mho}_{b,l-1}^2 \bm{b}_{m,l}, \\
		\bm{k}_{b,m,l} &=\eta_{b,e,m,l}^{-2} \bm{A}_{l} \bm{\mho}_{b,m,l-1}^2 \bm{b}_{m,l},  \\
		\bm{e}_{m+1,l} &=\bm{e}_{m,l}-\hat{\boldsymbol{H}}_{m,l-1}^{H}\bm{b}_{m,l}, \\
		\hat{\boldsymbol{H}}_{m,l+1} &=\bm{A}_{l} \hat{\boldsymbol{H}}_{m,l}+\bm{k}_{b,m,l} \bm{e}_{m+1,l}^H, \\
		\bm{\mho}_{b,m,l}^2 &=\bm{A}_{l} \bm{\mho}_{b,m,l-1}^2 \bm{A}_{l}^{H}+   \Gamma_{\Delta}^2 -\eta_{b,e,m,l}^2\bm{k}_{b,m,l}\bm{k}_{b,m,l}^{H}.	\nonumber	\end{aligned}
\end{align}
\end{algorithm}

%%%%%%%%%%%%%%%%%%%%%%%%%%%%%%%%%%%%%%%%%%%%%%%%%%%%%%%%%%%%%%%%%%%%%%%%%%%%%%%
%%%%%%%%%%%%%%%%%%%%%%%%%%%%%%%%%%%%%%%%%%%%%%%%%%%%%%%%%%%%%%%%%%%%%%%%%%%%%%%

%\clearpage\newpage
\section{Hysteresis Loop used in the Experiments as Nonlinearity} \label{app:B}
In acoustic reproduction systems, a magnet is used to convert the electrical signal into the motion of the loudspeaker diaphragm. Due to the magnetic hysteresis, the loudspeaker imposes a strong nonlinearity on the playback signal \cite{soria2004modeling}. This operation happens during transformation of an electrical signal into diaphragm displacement.
The Bouc-Wen model is used to simulate this hysteresis system, \cite{Bouc01, wen1976method}. The following two equations describe the state-space equations of the Bouc-Wen model where \eqref{eq:hys_state_update} is the state equation and \eqref{eq:hys_output_update} is the output equation.
\begin{align} \label{eq:hys_state_update}
\frac{ds(t)}{dt} &= \alpha \frac{du(t)}{dt} - \zeta |s(t)| \frac{du(t)}{dt} - \beta | \frac{du(t)}{dt} | s(t), \\
\label{eq:hys_output_update}
d(t) &= \mu u(t) - s(t),
\end{align}
where $u(t)$ is the input voltage, $d(t)$ is the output displacement, and $s(t)$ is the hysteretic state variable. $\alpha$, $\beta$, $\zeta$, and $\mu$ are the system parameters which determine the shape and the hysteresis loop. One can use the forward Euler method to find the discrete version of the state-space equations as 
\begin{align}
s[n] = & s[n-1] + \alpha \left ( u[n] - u[n-1] \right ) \\ \nonumber
& - \zeta \left |  s[n-1]  \right |  \left (  u[n] - u[n-1]  \right ) \\ \nonumber
& - \beta \left |   u[n]  - u[n-1]  \right | s[n-1], \\
d[n] = &\mu u[n] - s[n].
\end{align}

\begin{figure}[htp]
\centering
\includegraphics[width=0.45\textwidth]{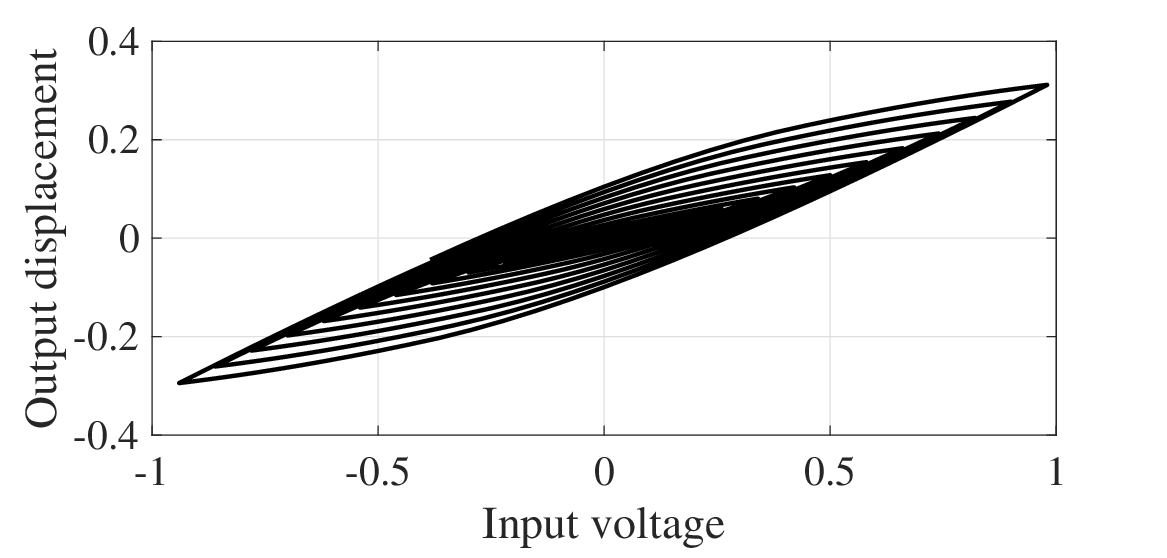}
\vspace{-2pt}
\caption{The hysteresis loop for $\alpha = 0.3$, $\beta = 1$, $\zeta = 0.5$, and $\mu= 0.5$.}
\label{fig:hys_loop}
\end{figure}

In order to create a strong nonlinearity, we set the model parameters $\alpha$ to $0.3$, $\beta$ to $1$, $\zeta$ to $0.5$, and $\mu$ to $0.5$. Figure \ref{fig:hys_loop} shows the hysteresis loop created by this model. 

%\clearpage\newpage
\section{Amplitude Modulation as Nonlinearity} \label{app:modulation}
\begin{figure}[htp]
\centering
\includegraphics[width=0.5\textwidth]{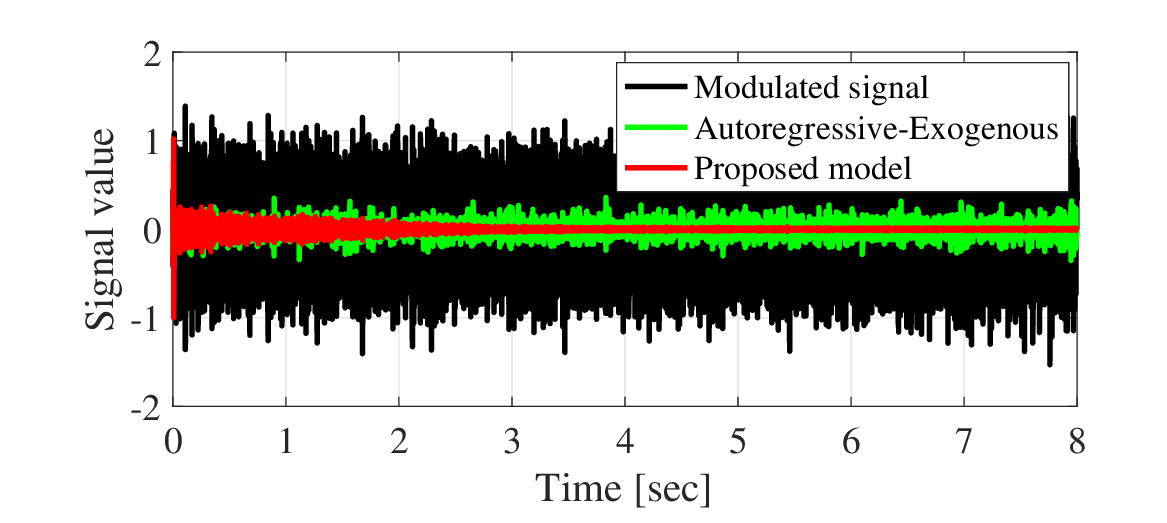}
%\vspace{-8pt}
\caption{This figure shows the output of a modulation system to a random noise and the error signal of the modeled system and Autoregressive-Exogenous. The total modeling error $\delta$ is $-32.96$ dB.}
\label{fig:modulation}
\end{figure}

Often, for the synthesis of typical audio signals, amplitude modulation is employed \cite{koumura2019cascaded}. In amplitude modulation, the input signal is multiplied by a sine wave. This shifts the spectrum of the input signal. 
In this experiment, we use a random noise as an input excitation of a modulation system. This system modulates the input signal by a $\pi/2$ [radians/sec] tone. 
\begin{align}
y[n] = x[n] \sin \left (  \frac{\pi}{2} n  \right ).
\end{align}
For Neural Harmonium the total number of bins $N_s$ is set to $32$, hop size $N_h$ to $16$, and the number of stages $M$ to $15$.
Again we compare our model to a nonlinear autoregressive exogenous model with a wavelet network (ARX). In this experiment, we deferred from presenting Wiener-Hammerstein model, and Gaussian process regression due to the observed  poor performance.
The ARX model achieved $-2.99$ dB modeling error which was outperformed by our proposed model achieving $-32.96$ dB. The modeling errors are shown in Figure \ref{fig:modulation}.

\begin{figure}[!h]
\centering
\includegraphics[width=0.35\textwidth]{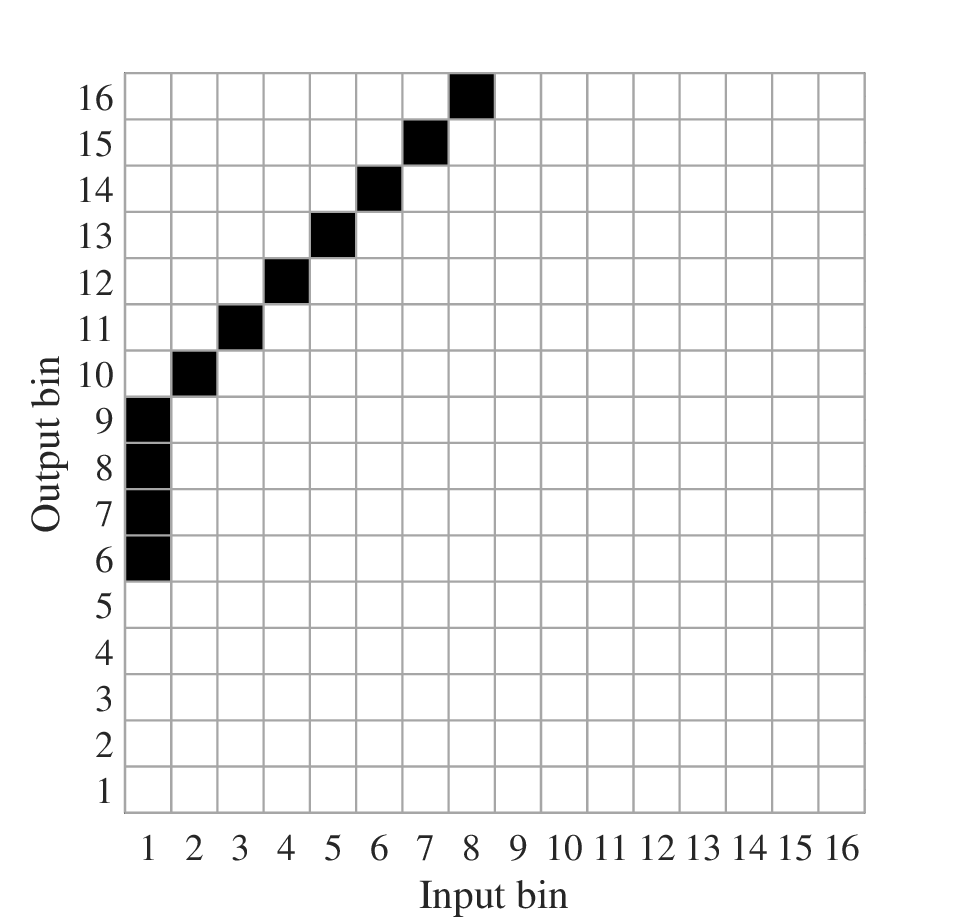}
\caption{The dependency map of the modulation model.}%\vspace{6cm}
\label{fig:hys_dependency_map_mod}
\end{figure}

Figure~\ref{fig:hys_dependency_map_mod} depicts the identified dependency map for this example. The off-diagonal structure of the resulting map portrays the spectral shift caused by the modulation. 

\end{document}